\documentstyle[11pt,aaspp4]{article}

\input{psfig.sty}

\begin{document}

\title{Evidence for a Sudden Magnetic Field Reconfiguration in SGR~1900+14}

\author{
Peter~M.~Woods\altaffilmark{1,2},
Chryssa~Kouveliotou\altaffilmark{1,2},
Ersin~{G\"o\u{g}\"u\c{s}}\altaffilmark{2,3},
Mark~H.~Finger\altaffilmark{1,2},
Jean~Swank\altaffilmark{4}, 
Don~A.~Smith\altaffilmark{5}, 
Kevin~Hurley\altaffilmark{6}, and
Christopher~Thompson\altaffilmark{7}
}

\altaffiltext{1}{Universities Space Research Association; 
peter.woods@msfc.nasa.gov}
\altaffiltext{2}{NASA Marshall Space Flight Center, SD50, Huntsville, AL
35812}
\altaffiltext{3}{Department of Physics, University of Alabama in Huntsville, 
Huntsville, AL 35899}
\altaffiltext{4}{NASA Goddard Space Flight Center, Greenbelt, MD 20771}
\altaffiltext{5}{Department of Physics and Center for Space Research, 
Massachusetts Institute of Technology, Cambridge, MA 02138}
\altaffiltext{6}{University of California, Berkeley, Space Sciences Laboratory,
Berkeley, CA 94720$-$7450}
\altaffiltext{7}{Canadian Institute for Theoretical Physics (CITA), 60 St.
George St., Toronto, M5S 3H8, Canada}

\begin{abstract}

We report the detection of large flux changes in the persistent X-ray flux of
SGR~1900$+$14 during its burst active episode in 1998.  Most notably, we find a
factor $\sim$700 increase in the non-burst X-ray flux following the August
27$^{\rm th}$ flare, which decayed in time as a power-law.  Our measurements
indicate that the pulse fraction remains constant throughout this decay.  This
suggests a global flux enhancement as a consequence of the August 27$^{\rm th}$
flare rather than localized heating.  While the persistent flux has since
recovered to the pre-outburst level, the pulse profile has not.  The pulse
shape changed to a near sinusoidal profile within the tail of the August
27$^{\rm th}$ flare (in $\gamma$-rays) and this effect has persisted for more
than 1.5 years (in X-rays).  The results presented here suggest the magnetic
field of the neutron star in SGR~1900$+$14 was significantly altered (perhaps
globally) {\it during} the giant flare of August 27.

\end{abstract}


\newpage

\section{Introduction}

Soft Gamma Repeaters (SGRs) are believed to be a rare class of magnetized
neutron stars whose collective behavior challenges physical models to explain
their observed properties.  There are currently four known SGRs and one
candidate source (see Hurley 2000 for a review).  They emit brief (durations
$\sim$0.1 s), intense (super-Eddington luminosities $\lesssim$10$^{42}$ ergs
s$^{-1}$) bursts of hard X-rays and soft $\gamma$-rays which repeat on
timescales of seconds to years.  On two exceptional occasions over the last 20
years, giant flares that reached $\sim$10$^{45}$ ergs s$^{-1}$ were observed. 
These two events were associated with SGR~0526$-$66 (Mazets et al.\ 1979) and
SGR~1900$+$14 (Hurley et al.\ 1999a).  Three of the four SGRs are positionally
coincident with young supernova remnants (SNRs; SGR~1900$+$14 lies 2$^{\prime}$
outside G42.0$+$0.8 and its association is more suspect [Vrba et al.\ 2000;
Lorimer \& Xilouris 2000]) and all are associated with persistent X-ray
counterparts.  These counterparts have luminosities in quiescence of
$\sim10^{34}~-~10^{35}$ ergs s$^{-1}$ and spectra that can be roughly
characterized by a simple power-law ($\alpha \sim-$2.2) attenuated by
interstellar absorption.  Coherent pulsations have recently been detected in
the persistent emission from two of these sources (SGR~1806$-$20 at 7.5 s
[Kouveliotou et al.\ 1998] and SGR~1900$+$14 at 5.2 s [Hurley et al.\ 1999b]). 
Both sources are spinning down rapidly at rates $\sim$10$^{-10}$ s s$^{-1}$
(Kouveliotou et al.\ 1998 and Kouveliotou et al.\ 1999, respectively).

Most physical models for SGRs agree that they are young, isolated, magnetized
neutron stars.  The theories differ on the proposed strength of the stellar
magnetic field, and hence, the energy source of the emitted radiation.  Models
that suggest the SGRs have field strengths $\sim10^{11}-10^{12}$ G (e.g.\
Marsden et al.\ 2000) requires accretion to account for the persistent X-ray
emission.  Such models have not yet offered any explanations for the
hyper-Eddington burst emissions.  On the other hand, Thompson \& Duncan (1995,
1996) have put forward a model for the SGRs as highly magnetized neutron stars
($\sim10^{14}-10^{15}$ G), i.e.\ magnetars, which can account for both the
persistent and burst emission properties.  In the magnetar model, the decay of
the superstrong magnetic field powers the persistent X-ray emission through
low-level seismic activity and heating of the stellar interior (Thompson \&
Duncan 1996).  The bursts are due to large-scale crust fractures that are
driven by the evolving magnetic field (Thompson \& Duncan 1995).  The
super-Eddington burst fluxes can be achieved in the presence of such a strong
field due to the suppression of the electron scattering cross-section for some
polarizations (Paczy\'nski 1992).

SGR~1900$+$14 was discovered after only three bursts were recorded from the
source in 1979 (Mazets \& Golenetskii 1981).  Thirteen years later, four more
events were detected from this SGR (Kouveliotou et al.\ 1993).  The pulsed
signal at 5.2 s from SGR~1900$+$14 was discovered (Hurley et al.\ 1999b) during
an April 1998 observation of the source with the {\it Advanced Satellite for
Cosmology and Astrophysics} (ASCA).  This observation coincidentally took place
just three weeks prior to burst reactivation of the SGR (Hurley et al.\
1999c).  Subsequent observations with the Proportional Counter Array (PCA)
aboard the {\it Rossi X-ray Timing Explorer} (RXTE) of SGR~1900$+$14 confirmed
the pulsations and established that the source was spinning down rapidly,
having a period derivative $\sim10^{-10}$ s s$^{-1}$ (Kouveliotou et al.\
1999).

The peak of the burst active phase for SGR~1900$+$14 was reached on 1998 August
27 when a giant flare was recorded by numerous instruments.  This flare started
with a short ($\sim$0.07 s), soft spike (often referred to as the
``pre-cursor'') that was followed by a much brighter, hard pulse (duration
$\sim$1 s) that approached $\sim$10$^{45}$ ergs s$^{-1}$, and a soft
$\gamma$-ray tail modulated at 5.2 s (Hurley et al.\ 1999a; Feroci et al.\
1999; Mazets et al.\ 1999).  The 5.2 s oscillating tail decayed in a
quasi-exponential manner over the next $\sim$6 min (Feroci et al.\ 2000). 
Integrating over the entire flare (assuming isotropic emission), at least
$\sim$10$^{44}$ ergs were released in $\gamma$-rays greater than 15 keV (Mazets
et al.\ 1999).  Shortly after this flare, the All-Sky Monitor (ASM) aboard RXTE
detected SGR~1900$+$14 for the first time in its four year mission at a flux
level of $\sim$0.1 Crab (Remillard, Smith \& Levine 1998).  Approximately one
week following the bright August 27$^{\rm th}$ flare, a transient radio flare
lasting $\sim$10 days was recorded with the Very Large Array (Frail, Kulkarni
\& Bloom 1999).  The temporal decay of this radio flare is consistent with a
power-law having an exponent of $-$2.6 $\pm$ 1.5 (Frail et al.\ 1999).  On
August 29, another bright SGR burst was detected which resembled the August
27$^{\rm th}$ flare in many ways (Ibrahim et al.\ 2000).  Like the giant flare,
this burst had a well-defined, relatively weaker pre-cursor and was followed by
a long ($\gtrsim$1000 s), oscillitory tail.  However, this burst was scaled
down in both peak luminosity and duration in $\gamma$-rays by a factor
$\sim$100.

Sixteen days after the August 27$^{\rm th}$ flare, imaged X-ray observations
with ASCA and the {\it Satellite per Astronomia X} (BeppoSAX) of the SGR
recorded the persistent flux level of the source, which had grown by a factor
$\sim$2.5 above the value prior to burst activity (Murakami et al.\ 1999; Woods
et al.\ 1999a).  From an earlier observation of SGR~1900$+$14 with BeppoSAX in
1997 May, while the source was in quiescence, we found that the spectrum could
not be fit by a simple power-law model; however, the sum of a blackbody and a
power-law provided an adequate fit to the 0.1$-$10 keV source spectrum (Woods
et al.\ 1999a).  A recent analysis of the ASCA observation in 1998 April, 
preceding reactivation by $\sim$3 weeks, shows that the two-component model
(blackbody $+$ power-law) yields a significantly better fit to the data
(D.~Marsden, private communication).  By including a low temperature ($kT$
$\sim$0.5 keV) blackbody component in the spectral model, the power-law index
becomes flatter to accomodate the blackbody flux contribution which is most
prominent at the low end of the observed energy range.  Consequently, the
photon index measured using the two-component model ($\alpha=1.11\pm0.19$ in
1997 May) differs significantly from the index as measured using a simple
power-law model for the same data ($\alpha=1.9$ [Woods et al.\ 1999a]).  When
the source became more luminous during the burst active phase, the X-ray energy
spectrum could be adequately modeled with a simple power-law.  The thermal
component was no longer significant, hence, the flux enhancement was attributed
to a rise in the power-law component of the spectrum (Woods et al.\ 1999a). 
Note that this does not imply the blackbody component ``disappeared'' or even
faded, but was rather overwhelmed by the much brighter power-law component.

Here, we have analyzed a large set of X-ray observations of the source in order
to construct a more complete flux history for SGR~1900$+$14.  We compare
changes observed in the pulse profile during the tail of the August 27$^{\rm
th}$ event with changes in the pulse profile of the persistent emission.  We
discuss these observations and the constraints they place on different models
for the SGRs.  Finally, we suggest that these new observations support the idea
of a long-lasting reconfiguration of the stellar magnetic field that took place
{\it during} the August 27$^{\rm th}$ flare.

\section{The X-ray Data Set}

Since 1992, there have been only 8 pointed observations with X-ray imaging
instruments of the persistent counterpart to SGR~1900$+$14 (not including
surveys).  Two observations were performed with the {\it Roentgen Satellite}
(ROSAT) High Resolution Imager (HRI), two with ASCA, and four with the BeppoSAX
Narrow Field Instruments (NFI).  To supplement this sparse data set, we have
included more than 20 other observations using the PCA and ASM aboard RXTE. 
The PCA observations were particularly well-sampled during the burst active
phase of the source in 1998.  This time period is of key interest since it
contains both the BeppoSAX and ASCA measurements of a $\sim$2.5 factor
intensity increase in the persistent flux above the quiescent level and the ASM
detection at 0.1 Crab.

With the exception of the two ROSAT HRI (0.1$-$2.4 keV) observations in 1994
and 1995, the instruments that observed SGR~1900$+$14 had good sensitivity
within the energy range 2$-$10 keV.  To make an adequate comparison of the
source flux between different epochs, we have chosen to use this nominal energy
range (2$-$10 keV) for flux measurements.  For this reason, we have excluded
the ROSAT observations from our sample.  For the first two BeppoSAX NFI
observations (1997 May and 1998 September) as well as for the two ASCA
observations, we used the flux measurements reported earlier (Woods et al.\
1999a, Hurley et al.\ 1999b and Murakami et al.\ 1999, respectively).  We
report here on more recent NFI observations made in March and April 2000.

\subsection{BeppoSAX NFI Observations}

The BeppoSAX NFI observed the source on 2000 March 30 and April 26,
approximately one year after the last detected burst activity from
SGR~1900$+$14.  The source exposure times were 40 ks and 40 ks for the
Medium-Energy Concentrator Spectrometers (MECS) and 15 ks and 18 ks for the
Low-Energy Concentrator Spectrometer (LECS) during the respective
observations.  In each observation, the SGR was aligned with the optical axis
of the instruments.  We used source extraction regions of 4$^{\prime}$ and
8$^{\prime}$ for the MECS and LECS, respectively.  Due to the low Galactic
latitude and dim intensity of SGR~1900$+$14, concentric rings of
6.4$^{\prime}-$9.6$^{\prime}$ and 9$^{\prime}-$13$^{\prime}$ were chosen from
each pointing for background subtraction with the MECS and LECS, respectively. 
The resulting spectra were then analyzed using XSPEC v10.00 along with the most
recent response matrices and effective area corrections.

We fit the data (0.12$-$10.5) to four different models, a simple power-law
(PL), a blackbody, a thermal bremsstrahlung, and a power-law plus a blackbody
(PL$+$BB), all modified by interstellar absorption.  Both the blackbody and
thermal bremsstrahlung models yielded poor fits to the data, while the PL and
PL$+$BB models returned reduced $\chi^2$ values below 2.  We find no
significant difference in the spectral parameters (including source flux)
between the two observations.  For this reason, we decided to fit the two data
sets simultaneously in order to better constrain the model parameters.  We did
allow for independent normalization factors for the LECS and MECS instruments
as there tends to be a 5$-$10\% systematic difference between the detector
normalizations.  As with the pre-outburst BeppoSAX observation from 1997 May,
the PL$+$BB model provided the best fit to the data (see Table 1).  We
quantified the preference of this model over the PL model using the F-test. 
The probability that the measured $\chi^2$ difference (31) between the PL$+$BB
model and the PL model would occur by chance is 1 $\times$ 10$^{-6}$.  The
unabsorbed 2$-$10 keV X-ray flux measured in 2000 March/April is 1.03(5)
$\times$ 10$^{-11}$ ergs cm$^{-2}$ s$^{-1}$ with $\sim$20\% of this flux coming
from the thermal blackbody component.  We note that the flux level in these
recent observations is consistent with the flux measured in 1997 May (F$_x$ =
0.99(4) $\times$ 10$^{-11}$ ergs cm$^{-2}$ s$^{-1}$ [Woods et al.\ 1999a]),
hence, we conclude that the source has returned to its quiescent flux level. 
Furthermore, all other spectral parameters measured in the 2000 observations
are consistent with the 1997 BeppoSAX measurements with the exception of the
power-law photon index ($\alpha$) which has steepened slightly (3.5$\sigma$)
between the two observations.

The event times for the combined MECS units were corrected to the Solar-system
barycenter and we then performed an epoch-fold period search within each
observation between 5.15 and 5.25 s.  We detect the pulsed signal in each
observation and measure significant spin down since the last reported X-ray
measurements in early 1999 (Woods et al.\ 1999b) as well as between the two
BeppoSAX observations.  The details of the timing of SGR~1900$+$14 will be
discussed elsewhere (Woods et al.\ 2001).  From the earlier BeppoSAX
observations, we noted that despite a large change in source flux and pulse
profile of the SGR, the root-mean-square (RMS) pulsed fraction remained
constant.  For the observations of 2000 March and April, we measure RMS pulse
fractions of 9.4\% $\pm$ 1.7\% and 10.4\% $\pm$ 1.7\%, respectively.  These
values are consistent with the measurements from 1998 September (11.4\% $\pm$
1.5\%) and 1997 May (12.2\% $\pm$ 1.1\%).  This result shows the pulse fraction
is constant at least within the four BeppoSAX observations which span $\sim$3
years and varying levels of source intensity and activity.

\subsection{ASM Observations}

Remillard et al.\ (1998) reported that analysis of the real-time data stream
from the RXTE All-Sky Monitor (ASM) revealed a detection of SGR~1900$+$14 at
$\sim$100 mCrab (2$-$12 keV) approximately two hours after the giant flare of
August 27.  A reexamination of the complete production light curve shows that
an additional 90-s observation was performed at only $\sim$24 minutes after the
flare, and this observation found SGR~1900$+$14 to be 270~$\pm$~17 mCrab.  By
$\sim$7 h after the flare, the intensity of the source had fallen below the ASM
detection threshhold of $\sim$20 mCrab, and has not been detected reliably
since.  An observation just $\sim$20 minutes before the flare shows no
detectable emission from SGR~1900$+$14.

The ASM performed two passes over SGR~1900$+$14 during the 2.5 h after the
flare.  We have checked the Burst and Transient Source Experiment (BATSE) data
for the five 90-s ASM observations performed during these passes.  We find that
the SGR was not occulted by the Earth for BATSE and there is no indication of
significant burst activity in the large-area detector discriminator data (1.024
s time resolution).  Therefore, the following ASM measurements reflect only the
persistent, non-burst flux level.  During each of the five 90-s ASM
observations, or dwells, the best-fit average count rates for the SGR were
calculated in three energy channels (1.5$-$3, 3$-$5, and 5$-$12 keV), and these
count rates were averaged over each pass to yield two intensity measurements. 
To convert these intensities into fluxes, we took a simple power-law spectrum
with a fixed slope of $-$2.2, as derived from the BeppoSAX observations of 1998
September 15, integrated over an estimated effective area for the ASM, and fit
the spectral normalizations to the observed intensties.  The effective area
table for the ASM is a theoretical construct, compiled during the design phase
of the experiment, and may differ from the actual response of the instrument. 
The unabsorbed flux (2$-$10 keV) at 0.4 and 2.1 hours following the beginning
of the flare decayed from (7.2~$\pm$~0.5) to (2.6~$\pm$~0.5) $\times$ 10$^{-9}$
ergs cm$^{-2}$ s$^{-1}$.  The errors here represent the statistical errors
convolved with a 3\% systematic error as is standard for the ASM data.  There
are likely further systematic errors that we are not able to accurately account
for (e.g.\ uncertainties in the ASM effective area, spectral evolution between
the ASM observations and the BeppoSAX observations).

\subsection{PCA Observations}

Due to the fact that the PCA is not an imaging instrument and there exist at
least two bright and variable X-ray sources (4U~1907$+$09 and XTE~J1906$+$09)
close in angle to SGR~1900$+$14, extracting the SGR flux from these
measurements required an alternative approach.  From the four BeppoSAX NFI
observations of SGR~1900$+$14, we know that despite the observed changes in
source intensity and pulse profile, the RMS pulsed fraction remained constant. 
If this property of the SGR holds at all times, then a measure of the pulsed
intensity would relate directly to the net source intensity.  A pulsed
intensity measurement with the PCA is much ``cleaner'' in that there is no
ambiguity with where the signal originates.  Under the premise that the pulsed
fraction remains constant, we set out to measure the RMS pulsed intensity of
SGR~1900$+$14 and indirectly, the net source flux using the PCA observations.

First, the data were grouped into segments where the PCA instrument
configuration was constant, that is, observations with the same set of PCUs on
and a fixed pointing.  One $\sim$3 day segment near 1998 September 8 was
eliminated from our sample due to the occurrence of a bright outburst from
XTE~J1906$+$09 (Takeshima, Corbet \& Swank 1998) which dominated the count rate
in the PCA.  Next, the 2$-$10 keV data for each selected interval were folded
on the phase connected solutions reported in Woods et al.\ (1999b) for the data
prior to 1999 and on ephemerides reported in Woods et al.\ (2001) for
subsequent observations.  For each detector configuration, the collimator
efficiency\footnote{The collimator efficiency files for each PCU can be found
at the following www address:
http://lheawww.gsfc.nasa.gov/users/keith/fieldofview/collimator.html} in units
of counts s$^{-1}$ PCU3$^{-1}$ was calculated and the count rates in each phase
bin of the folded profile were corrected accordingly.  Finally, the RMS pulsed
intensity for each folded light curve was measured.

\section{The Flux History}

\begin{figure}[!p]
\centerline{
\psfig{file=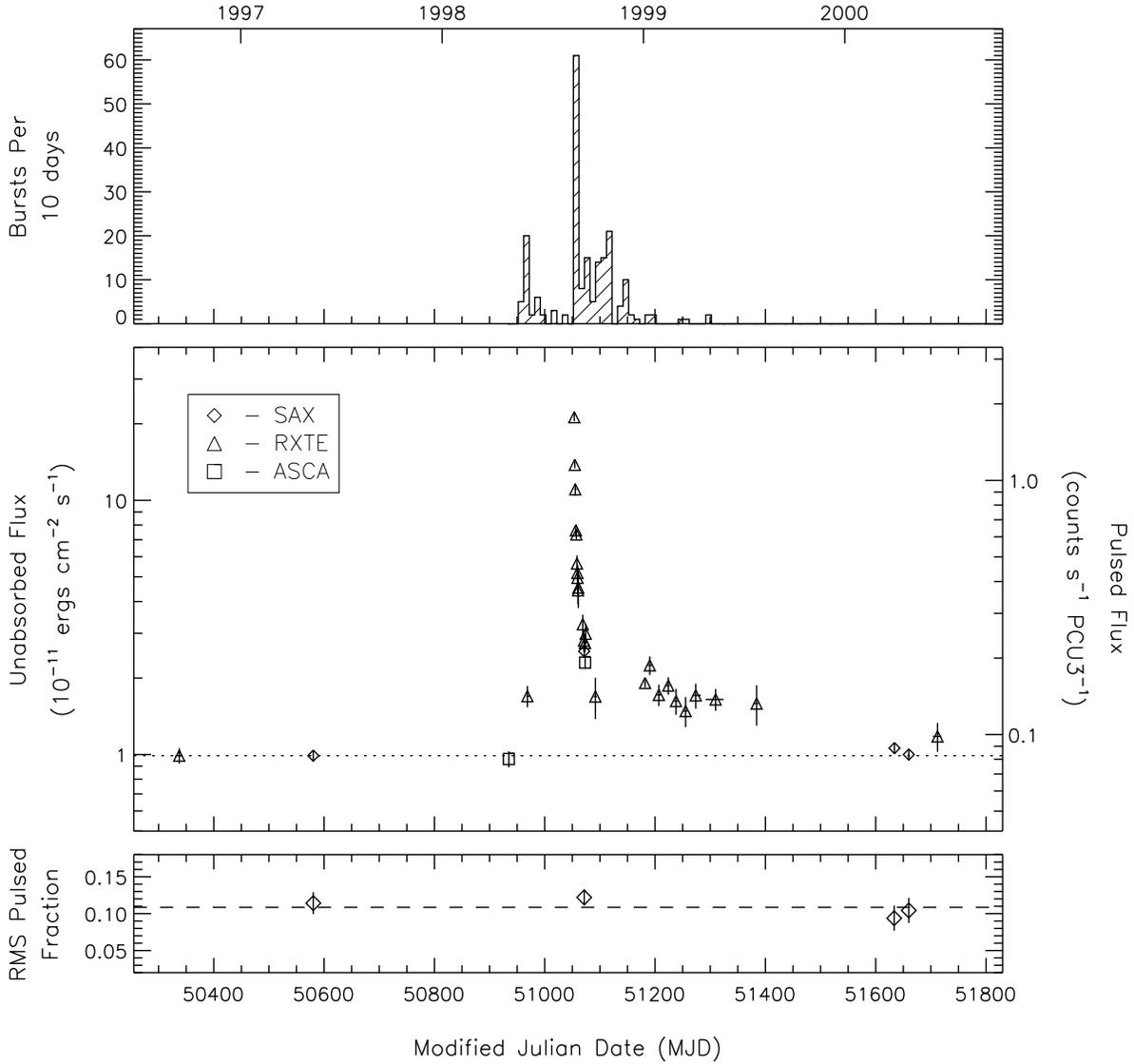,height=6.5in}}
\vspace{-0.15in}

\caption{{\it Top panel} -- Burst rate history of SGR~1900$+$14 as observed
with BATSE.  {\it Middle panel} -- Flux history of SGR~1900$+$14 covering 3.8
years.  The left vertical scale is unabsorbed 2$-$10 keV flux and the right is
pulsed flux in units of counts s$^{-1}$ PCU3$^{-1}$.  The dotted line marks the
flux level as observed by BeppoSAX in 1997 May which was used for normalizing
the pulsed flux measurements.  See text for further details.  {\it Bottom
panel} -- Pulse fraction of SGR~1900$+$14 (2$-$10 keV) as measured within the
four BeppoSAX observations using the MECS instruments.  The dashed line marks
the mean RMS pulsed fraction ($f_{\rm RMS} \sim$ 0.11).}

\vspace{11pt}
\end{figure}

The flux history of SGR~1900$+$14 over $\sim$3.8 years is given in the middle
panel of Figure 1.  The scale on the left vertical axis corresponds to the
BeppoSAX and ASCA measurements.  The right axis applies to the PCA pulsed flux
measurements.  The relative normalization between the pulsed flux measurements
and the net source flux measurements was calculated in the following way. 
Prior to burst reactivation of the source in 1998 May, there was one
observation each with RXTE, BeppoSAX, and ASCA.  The two flux measurements
during quiescence with the two imaging instruments, BeppoSAX in 1997 May and
ASCA in 1998 April, are consistent with one another.  Since our only
measurements of a change in source flux are during a burst active interval, we
assume the SGR flux is constant in quiescence.  Therefore, we defined the
normalization to be the ratio of the source flux as measured by BeppoSAX prior
to burst activity in 1997 May to the PCA pulsed intensity measurement from 1996
September.  We use this normalization to define the source flux from the PCA
measurements during the burst active period.  Note the good agreement of the
pulsed intensity measurements (PCA) to nearly simultaneous net source intensity
measurements (BeppoSAX and ASCA) following reactivation (inset of Figure 2). 
This strengthens our initial hypothesis that the pulsed fraction remains
constant, even during burst active phases of the source.  The pulsed intensity
observations made with the PCA suggest the net source intensity of
SGR~1900$+$14 increased by a factor $\sim$20 over the quiescent level
approximately 1 day after the giant flare.

We next compared the changes observed in source flux with the change in burst
activity.  We quantified the burst activity using the burst rate as observed
with BATSE (top panel of Figure 1).  Note the good correlation between the rise
and decay of the burst activity and the rise and decay of the SGR flux.  We
find a significant increase of the source flux during the initial burst
activity of 1998 May/June, which shows the flux enhancement correlates not only
with the giant flare, but also with the more common recurrent bursts.  The next
measurements were made starting $\sim$1 day after the giant flare and found the
SGR flux was more than an order of magnitude brighter than the level during
quiescence.  This transient flux enhancement was an artifact of the August
27$^{\rm th}$ flare, lasting $\sim$40 days, and is discussed in greater detail
below.  The next sequence of pulsed flux measurements was acquired over the
first seven months of 1999.  During these observations, we observe a gradual
decline of the X-ray flux as well as the burst occurrence rate of the SGR.  It
is not clear from these observations whether the flux enhancement within this
portion of the X-ray lightcurve is connected with a slow decay component
associated with the August 27$^{\rm th}$ flare (Kouveliotou et al.\ 2001), or
instead, the sum of multiple smaller enhancements from burst activity in the
last few months of 1998 and the first half of 1999.  The next observations of
SGR~1900$+$14 were performed in 2000 March$-$July, starting 11 months after the
last recorded burst emission from the source.  These observations found the
source flux had returned to the pre-outburst level (see also section 2).

Focusing now on the large flux change observed following the giant flare, we
find the net/pulsed flux decayed according to a power-law relative to the onset
of the flare.  Using both the net and the scaled pulsed flux measurements, we
fit the data to a power-law and find a decay constant of $-$0.713 $\pm$ 0.025. 
This value is consistent with the power-law flux decay (F $\propto$ t$^{-0.8
\pm 0.1}$) found following the burst of August 29 (Ibrahim et al.\ 2000). 
Extrapolating the fit to the August 27$^{\rm th}$ X-ray light curve back
towards the flare itself, we find fair agreement between the flux level we
would expect and the ASM flux measurements (Figure 2).  We note, however, that
each of the ASM measurements are significantly higher than the expectation from
the fit.  This discrepancy may be the result of assumptions made on the source
spectrum and/or the instrumental response (see section 2.2), non-negligible
flux enhancements following the numerous bursts during late September, a second
component related to the flare itself, and/or a slightly later reference
epoch.  We know from the 1998 May/June PCA measurement that the recurrent burst
activity increased the persistent SGR flux by a factor $\sim$2.  The bursts
following the giant flare in late August and early September do not likely
contribute significantly to the already large enhancement present from the
August 27$^{\rm th}$ flare, however, the cumulative burst activity in mid/late
September may provide comparable flux enhancements.  This effect would tend to
flatten the observed power-law decay slope.  Another possibility is there are
two components present in the ASM measurements directly after the flare; one
associated with the long X-ray tail we observe for many days after the flare
and one directly connected to the flare itself.  Finally, the ASM measurements
move closer to the fit when we push forward our reference epoch to $\sim$14 min
after the {\it onset} of the flare.  The observed duration of the August
27$^{\rm th}$ flare in $\gamma$-rays ($>$15 keV) is $\sim$6 min, so it is
certainly conceivable that the SGR ``afterglow'' did not begin until somewhere
near the end of the flare.

\begin{figure}[!htb]
\centerline{
\psfig{file=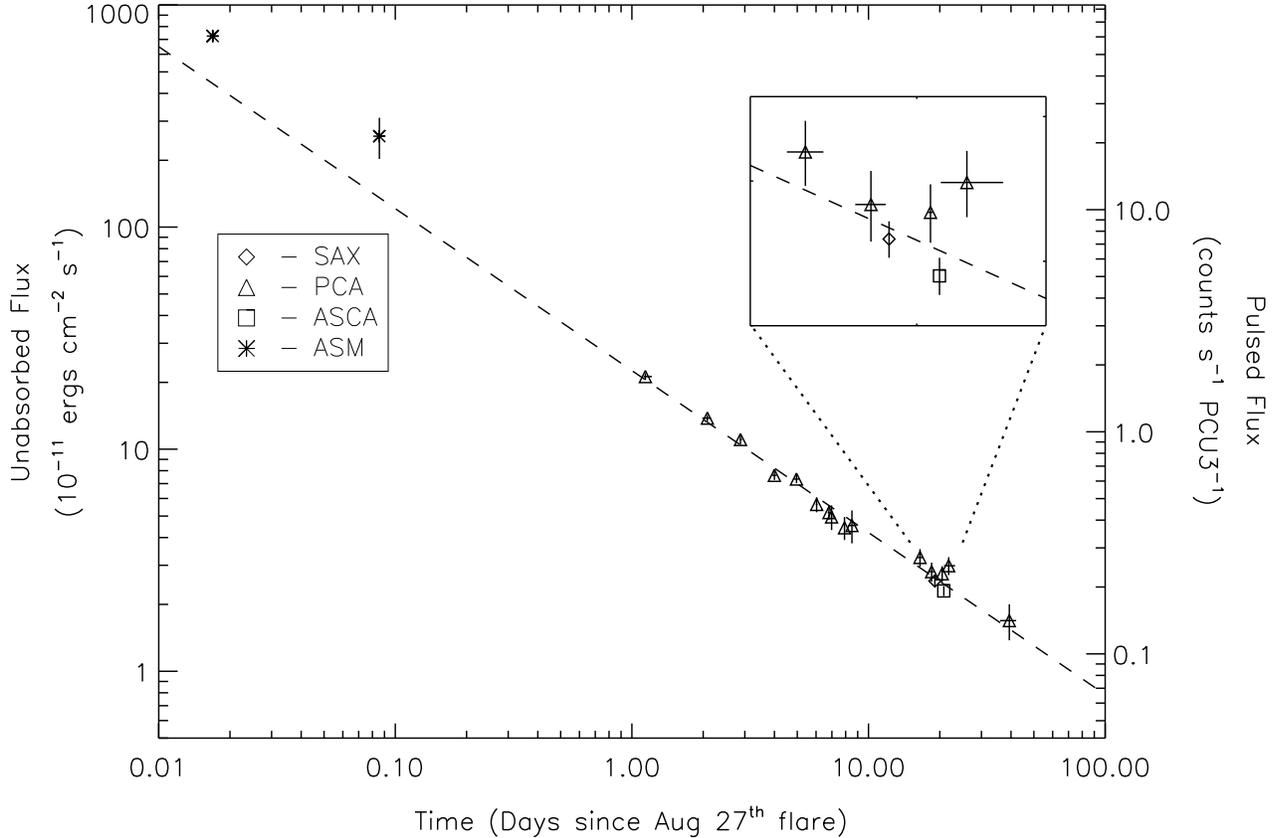,height=4.8in}}

\caption{Log-log plot of SGR~1900$+$14 flux versus time following the August
27$^{\rm th}$ flare.  The reference time is the beginning of the flare as
observed in soft $\gamma$-rays.  The dotted line is a fit to the RXTE/PCA,
BeppoSAX, and ASCA data only (i.e.\ the ASM data are not included in the fit). 
The slope of this line is $-$0.713 $\pm$ 0.025.  Inset shows the good agreement
between the pulsed flux measurements with the PCA and the absolute flux
measurements made with BeppoSAX and ASCA.}

\vspace{11pt}
\end{figure}

\section{Pulse Profile Changes}

In addition to undergoing large changes in source intensity, the pulse profile
of SGR~1900$+$14 was altered dramatically during the burst active interval of
1998.  Variation of the folded profile of the persistent X-ray source (2$-$10
keV) had been noted previously (Kouveliotou et al.\ 1999; Murakami et al.\
1999; Woods et al.\ 1999b) as well as the pulse-to-pulse evolution ($>$15 keV)
found within the tail of the August 27$^{\rm th}$ flare (Mazets et al.\ 1999;
Feroci et al.\ 2000).  Here, we combine the observations we have analyzed with
the archival data in order to show the evolution of the folded profile over the
last 3.8 years (Figure 3).  The light curves in the top row were all recorded
prior to the August 27$^{\rm th}$ flare while the panels in the bottom row are
all following.  Both top and bottom rows are folded over the energy range
2$-$10 keV.  The last panel on the top row was taken during/after significant
burst activity in 1998 May/June, yet it shows very little change from previous
observations.  The two middle panels were generated from Ulysses data (25$-$150
keV) of the giant flare.  The left panel was created by folding the data from
40$-$100 s after the onset of the flare, while the right panel displays the
folded profile from 280$-$330 s post-trigger.  The time resolution of the
Ulysses data is 0.5 s, therefore, these folded profiles are under-resolved. 
There are certainly finer structures in these profiles that are artificially
removed due to this effect.  The difference between the two light curves,
however, is evident: the complexity of the pulse profile diminishes with time
through the burst.  This is in agreement with what has been reported elsewhere
(Mazets et al.\ 1999; Feroci et al.\ 2000).  The same qualitative behavior can
be seen in the persistent emission by comparing the top and bottom rows of
Figure 3, despite being generated over a different energy range (2$-$10 keV). 
From this observation, we conclude that the bulk of the pulse profile change
observed in the persistent emission is {\it independent} of source flux and
appears to depend solely on the temporal relation of the observations to the
August 27${\rm th}$ flare.  This suggests that whatever caused the evolution of
the pulse profile through the giant flare, also produced the change in the
persistent X-ray light curve.

\begin{figure}[!p]
\centerline{
\psfig{file=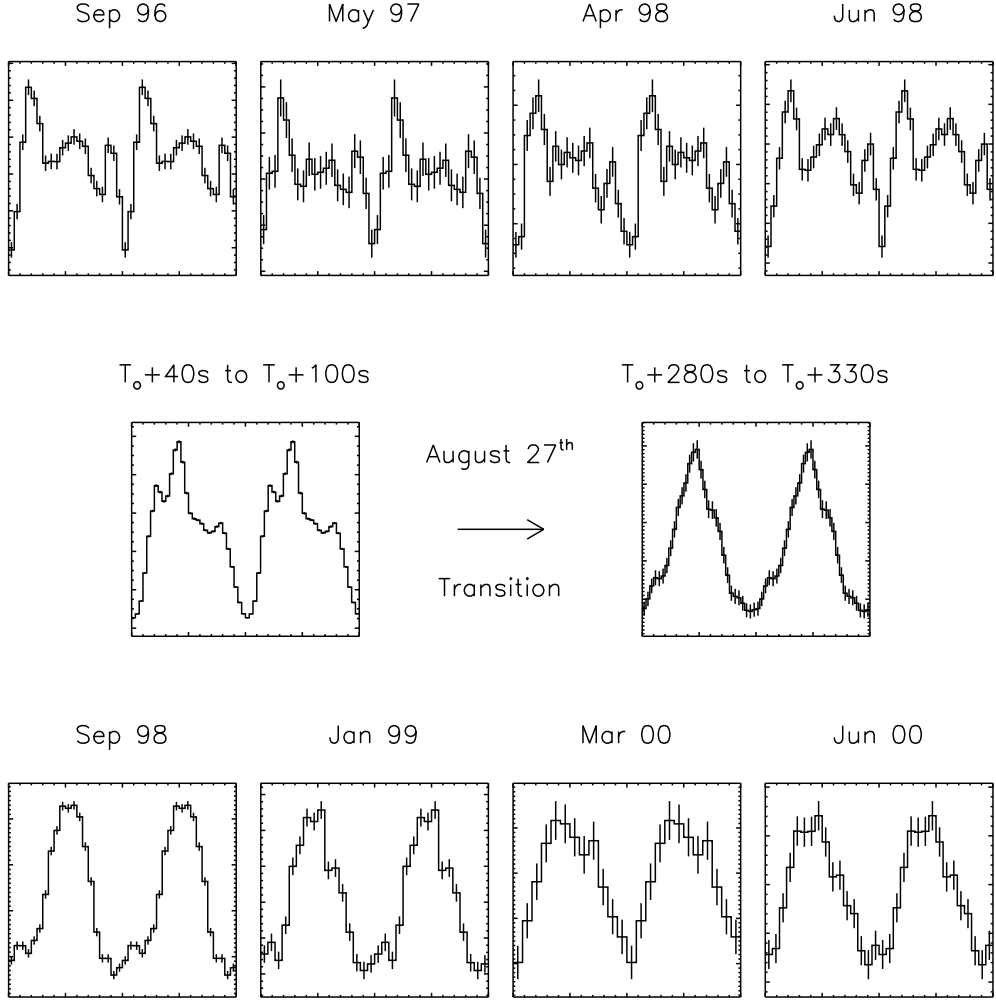,height=7.0in}}
\vspace{-0.7in}

\caption{Evolution of the pulse profile of SGR~1900$+$14 over the last 3.8
years.  All panels display two pulse cycles and the vertical axes are count
rates with arbitrary units.  The two middle panels were selected from Ulysses
data (25$-$150 keV) of the August 27$^{\rm th}$ flare.  Times over which the
Ulysses data were folded are given relative to the onset of the flare (T$_{\rm
o}$).  See text for further details.  The top and bottom rows are integrated
over the energy range 2$-$10 keV.  From top-to-bottom, left-to-right, the data
were recorded with the RXTE, BeppoSAX, ASCA, RXTE, RXTE, RXTE, BeppoSAX, and
RXTE.}

\end{figure}

\section{Discussion}

We have shown that the X-ray counterpart to SGR~1900$+$14 underwent large
changes in source intensity correlated with the burst activity of the source
during 1998.  Following the August 27${\rm th}$ flare, the persistent SGR flux
reached a maximum.  This flux enhancement decayed over the next $\sim$40 days
as a power-law in time with an exponent $-$0.71.  All measurements of the RMS
pulse fraction suggest this parameter remains constant despite the large
excursion in the X-ray intensity of the SGR.  Finally, we found the dramatic
pulse profile change is independent of the change in luminosity of the source,
but appears to be a direct consequence of the August 27$^{\rm th}$ flare.

Pulse profile changes in a fixed energy band are not uncommon in accreting
X-ray binaries.  However, these changes are typically correlated with
variations in mass accretion rate (see e.g.\ White, Nagase \& Parmar 1995). 
For a constant mass accretion rate (i.e.\ X-ray luminosity), the pulse profiles
tend to remain unchanged.  This is not the case for SGR~1900$+$14.  The
persistent source flux has now (March$-$July 2000) recovered to its quiescent
level, yet the pulse profile has remained invariant since 1998 August 28.  This
result severely constrains all accretion models for SGRs.

The giant flare of August 27 released a large amount of energy in $\gamma$-rays
($E_{\rm Aug~27} \sim 10^{44}$ ergs [Mazets et al.\ 1999]) within a relatively
short time period ($\sim$6 min).  A fraction of a percent of this energy was
stored by the star, and slowly re-radiated over the next $\sim$40 days to form
the observed power-law decay of the persistent X-ray flux (Figure 2).  Given
the high luminosity associated with the giant flare, the radiative momentum
released in the burst would excavate any hypothetical accretion disk out to a
large radius where the diffusion time is of the order of months to years
(Thompson et al. 2000).  Therefore, accretion could not be re-established in
such a short time ($\sim$10$^3$ s) as to account for the giant flare afterglow.
Furthermore, kinetic energy loss (rotation) is insufficient by orders of
magnitude as an energy source.  We conclude that the observed power-law flux
decay is clear evidence for a fundamentally new type of energy dissipation in
neutron stars.  

The constant pulse fraction through the long X-ray tail, and the non-recovery
of the pulse profile, argues strongly against the possibility of the afterglow
from the August 27$^{\rm th}$ flare being a strictly localized phenomenon.  If
the persistent, declining output were concentrated in a small fraction of the
stellar surface or magnetosphere without reprocessing at larger radius, then
the pulse fraction would be expected to change dramatically as the persistent
flux returned to its baseline value.   Taken at face value, this suggests that
energy was released globally throughout the crust and magnetosphere of the
star. The magnetosphere is the more likely location for the enhanced persistent
emission because the BeppoSAX observations show the flux increase occured in
the non-thermal power-law component of the persistent spectrum, not in the
thermal blackbody. 

We now address how the observed pulse profile change can be accommodated within
the framework of the magnetar model.  In the case of an isolated magnetar, the
pulse shape of the persistent X-ray source is governed by the distribution of
magnetic field and magnetospheric currents.  Since the flux has recovered to
the pre-outburst (i.e.\ quiescent) level, but the pulse profile has not, we
conclude that either the X-ray flux is being reprocessed far outside the region
where the non-thermal continuum is generated, or the distribution of currents
within the magnetosphere was severely altered {\it during} the August 27$^{\rm
th}$ flare.

There is good evidence that the surface magnetic field of SGR~1900$+$14 had a
multi-polar structure during the August 27$^{\rm th}$ giant flare, based upon
the complex profile observed in the large amplitude pulsations (Feroci et al.\
2000).  The hyper-Eddington flux of X-rays is easily  channeled along
(partially) open magnetic field lines in the presence of  a small amount of
matter.  The reduction of the pulse profile from four sub-pulses to a single
sub-pulse at the end of the giant flare has a simple explanation in terms of
the diminishing volume of a trapped fireball (Feroci et al.\ 2000).  Because
the persistent X-ray flux is channeled passively along the field near the star,
the higher multipoles are expected to remain following the giant flare. 
However, a giant flare may involve a global twist of the stellar magnetic
field, which drives a long-lived electrical current to re-scatter X-rays at the
electron cyclotron resonance near 10 stellar radii.  In this model, the
transient decay of the persistent emission requires the added presence of
sharper field gradients, which could be created by a fracture of the rigid
crust during the giant flare (Thompson et al.\ 2000).  In light of these new
observations, relevant concerns regarding this model would be the reappearance
of the blackbody (presumably surface) emission in the presence of the optically
thick scattering screen (which limits the amount of dissipation taking place
within the screen), as well as the constancy of the pulse fraction before and
after the giant flare.

We now consider an alternative explanation for the observed flux and pulse
profile changes in which the magnetospheric currents of SGR~1900$+$14 are
modified substantially through a global reconfiguration of the magnetic field. 
The reemergence of the blackbody component following the decay of the flux
enhancement suggests that we are again observing radiation directly from the
stellar surface.  The observation of the simple pulse profile in the presence
of the thermal component implies that the stellar magnetic field is now
predominantly dipolar.  Since the pulse profile was observed to change from
complex (multi-polar field structure) to simple (di-polar field structure)
within the August 27$^{\rm th}$ flare, this line of reasoning suggests that the
stellar magnetic field underwent a {\it global} reconfiguration at the time of
the giant flare.  This scenario requires a reorganization of the magnetic field
both inside and outside the star, which would proceed on a hydromagnetic
timescale rather than the much shorter dynamical time of the stellar
magnetosphere, and could plausibly take as long as $\sim$300 s.  A model
suggesting a global field reconfiguration during the giant flare was proposed
independently by Ioka (2000).  Ioka's model was developed to explain earlier
observations related to the spin period evolution of this SGR near the time of
the August 27$^{\rm th}$ flare (Woods et al.\ 1999b).  In Ioka's model, 
gravitational energy was proposed to power the flare through a substantial
change in the moment of inertia of the star and consequently, the total amount
of {\it magnetic} energy released ($\sim10^{49}$ ergs) would exceed by several
orders of magnitude the $\sim10^{44}$ ergs observed in the $\gamma$-ray band. 
One may also consider a global field reconfiguration scenario with magnetic
energy as the power source for the giant flare.  If the higher multipoles carry
a significant fraction of the external magnetic energy, then removing them
would require expending some 100$-$1000 times more energy than the observed
output of the flare.  In any model involving a global field reconfiguration of
a magnetar, the total internal/external field energy {\it must} be nearly
conserved during the transition, and currently, we find no compelling reason
for this to be the case.  In addition, if the magnetospheric currents were
significantly modified during the giant flare, then the near equality between
the ``baseline'' X-ray flux before and after the giant flare would be
coincidental.

A strong, evolving magnetic field is the foundation of the magnetar model for
the SGRs.  The dramatic change in the pulse profile of SGR~1900$+$14 following
the August 27$^{\rm th}$ flare provides strong evidence for some type of
large-scale reconfiguration of the magnetic field.  The two models involving
stellar field dynamics outlined above each have their own merits when compared
to the observations, however, both involve some level of coincidence. 
Moreover, the global field reconfiguration model posesses a fundamental flaw in
that the energetics of the model do not match the observations.  Without a
comprehensive model to explain the cause and long-lasting effects of the August
27$^{\rm th}$ flare, the giant flares, and to a lesser extent SGRs in general,
remain quite enigmatic.

\acknowledgments{\noindent {\it Acknowledgements} -- We thank the referee for
useful comments on this manuscript.  We are deeply indebted to the help we
received from the RXTE team, particularly Evan Smith and the SOF for scheduling
these extensive observations and Craig Markwardt for his assistance with part
of the data analysis.  We also thank the RXTE/SDC, the SAX/SDC (particularly
J.M.\ Muller), and HEASARC for pre-processing the RXTE/PCA and BeppoSAX data. 
We thank Rob Duncan for useful comments on this manuscript.  This work was
funded primarily through a Long Term Space Astrophysics program (NAG 5-9350)
for both PMW and CK.  MHF and EG acknowledge support from the cooperative
agreement NCC 8-65.  CT acknowledges support from the Alfred P. Sloan
Foundation and NASA grant NAG 5-3100.  PMW, CK, EG and CT appreciate useful
discussions at the ITP funded by NSF grant PHY99-07949. }

\newpage

\begin{center}
\begin{deluxetable}{cccccc}
\scriptsize
\tablecaption{Summary of spectral fits. \label{tbl-1}}
\tablewidth{6.5in}

\tablehead{
\colhead{Model}    &  
\colhead{N$_{\rm H}$ (10$^{22}$~cm$^{-2}$)}   &  \colhead{$kT$ (keV)}     &  
\colhead{R$_{\rm bb}$ (km)\tablenotemark{a}}  &  
\colhead{$\alpha$\tablenotemark{b}}          &
\colhead{$\chi^2$/dof} 
}

\startdata

BB$+$PL    &     2.7~$\pm$~0.4       &
	0.43~$\pm$~0.05  &  4~$\pm$~1    &  1.98~$\pm$~0.16     &  
	116.1/107 \nl

PL         &     2.4~$\pm$~0.2       &  
	...             &  ...              &  2.45~$\pm$~0.06   &
	146.8/110 \nl

\enddata

\tablenotetext{a}{Blackbody radius without general
relativistic correction (assumes d = 10 kpc) }
\tablenotetext{b}{Power law photon index}

\end{deluxetable}
\end{center}

\end{document}